\pdfoutput=1
\documentclass[useAMS,usenatbib]{mn2e}
\usepackage[pdftex]{graphicx}
\usepackage{times}
\usepackage{wasysym}
\usepackage{url}
\usepackage{lineno}
\usepackage{amssymb}

\newcommand{\g}{$\gamma$}

\newcommand{\HMS}[3]{$#1^{\mathrm{h}}#2^{\mathrm{m}}#3^{\mathrm{s}}$}
\newcommand{\DMS}[3]{$#1^\circ #2' #3''$}
\newcommand{\flux}{\,ph\,cm$^{-2}$\,s$^{-1}$}
\newcommand{\lumi}{\,erg\,s$^{-1}$}
\newcommand{\psr}{PSR~J1648$-$4611}
\newcommand{\hessj}{HESS~J1646$-$458}
\newcommand{\src}{FGL~J1651.6$-$4621}

\title{$\gamma$-ray emission from the Westerlund~1 region}

\author[S. Ohm, J.A. Hinton and R. White]{\normalsize S.~Ohm,$^{1}$, J.A.~Hinton$^{1}$ and R. White$^{1}$
  \\
  $^{1}$
  Department of Physics and Astronomy, The University of Leicester, University Road, Leicester, LE1 7RH, United Kingdom \\
}

\begin{document}

\date{Accepted 2013 June 24. Received 2013 June 21; in original form
  2013 May 02}
\pagerange{\pageref{firstpage}--\pageref{lastpage}}
\pubyear{2013}

\maketitle

\label{firstpage}

\begin{abstract}
  Westerlund 1 (Wd~1) is the most massive stellar cluster in the
  Galaxy and associated with an extended region of TeV emission. Here
  we report the results of a search for GeV $\gamma$-ray emission in
  this region. The analysis is based on $\sim$4.5 years of {\it
    Fermi}-LAT data and reveals significantly extended emission which
  we model as a Gaussian, resulting in a best-fit sigma of $\sigma_{S}
  = (0.475 \pm 0.05)^\circ$ and an offset from Wd~1 of
  $\sim$1$^\circ$. A partial overlap of the GeV emission with the TeV
  signal as reported by H.E.S.S. is found. We investigate the spectral
  and morphological characteristics of the $\gamma$-ray emission and
  discuss its origin in the context of two distinct
  scenarios. Acceleration of electrons in a Pulsar Wind Nebula
  provides a reasonably natural interpretation of the GeV emission,
  but leaves the TeV emission unexplained. A scenario in which protons
  are accelerated in or near Wd~1 in supernova explosion(s) and are
  diffusing away and interacting with molecular material, seems
  consistent with the observed GeV and TeV emission, but requires a
  very high energy input in protons, $\sim$$10^{51}$ erg, and rather
  slow diffusion.  Observations of Wd~1 with a future $\gamma$-ray
  detector such as CTA provide a very promising route to fully resolve
  the origin of the TeV and GeV emission in Wd~1 and provide a deeper
  understanding of the high-energy (HE) astrophysics of massive
  stellar clusters.
\end{abstract}


\maketitle

\section{Introduction}\label{sec:intro}

Supernova remnants (SNRs) have long been suggested as the dominant
source of Galactic cosmic rays (GCRs). The massive progenitor stars of
these supernova (SN) explosions are usually bound in associations or
stellar clusters and shape their immediate surroundings with their
fast supersonic winds or in interactions with the winds and/or
shockwaves of already exploded member stars. In this way a {\it
  superbubble}, filled with a hot tenuous plasma, can form. In such a
system particles may be accelerated by supersonic turbulences and/or
via repeated diffusive shock acceleration to TeV energies and
beyond~\citep[e.g.][]{Bykov01,Parizot04}. $\gamma$-ray observations from
hundreds of MeV to multi-TeV energies with satellite instruments and
using ground-based Cherenkov telescopes are an ideal tool to study not
only the acceleration sites of GCRs, but also their interaction with
and transport in the surrounding interstellar medium (ISM). Stellar
clusters and superbubbles are emerging as a new source population in
the $\gamma$-ray band. Associations include the non-thermal emission from
the Cygnus region \citep{HEGRA:2032, Fermi:Cygnus} and the TeV
emission from the young massive stellar clusters Westerlund~2 and
Westerlund~1 (Wd~1) \citep{HESS:Wd2,HESS:Wd2_2,HESS:Wd1_12}. These
observations suggest that the collective effect of stellar winds
and/or past SN activity in these complexes indeed results in a
significant production of GCRs.

Wd~1 is the most massive stellar cluster in the Milky Way with a total
mass between $\sim$5$\times 10^4 - 10^5\,M_\odot$ \citep{Clark05,
  Lim13} and a very rich population of post main sequence stars
\citep[e.g.][]{Clark02}. Distance estimates over the past ten years
seem to converge to $4.0-5.0$\,kpc, and we will employ the
$\sim$4.3\,kpc as used by \citet{HESS:Wd1_12} in the following. The
age estimate of the stellar cluster of $\sim $4\,Myr suggests that
$\sim$100 stars, all with progenitor masses $>$30\,$M_{\odot}$, could
have undergone SNe in Wd\,1, at an average rate of one per $10^{4}$
years \citep{Muno06b} over the past million years. The existence of a
magnetar in the cluster, with a progenitor mass presumably in excess
of $40\,M_\odot$ seems to confirm this estimate \citep{Muno06a}.

\citet{HESS:Wd1_12} reported on the detection of degree-scale
very-high-energy (VHE; $E > 100$\,GeV) emission from the vicinity of
Wd~1 with a total luminosity of $L_{\rm TeV} \simeq$2$\,\times 10^{35}
(d/4.3\,{\rm kpc})^{-2}$\lumi, representing a fraction of
$\sim$10$^{-4}$ of the total mechanical wind and SN power. Based on
the morphological and spectral properties of the H.E.S.S. source
(\hessj) \citet{HESS:Wd1_12} conclude that a significant part of the
TeV emission arises from proton-proton interactions of cosmic rays
(CRs) accelerated in and around Wd~1 that interact with ambient
material. In this scenario an overlap of the VHE $\gamma$-ray emission
with gas as traced in 21~cm line emission and in CO lines is expected
and indeed observed. In a proton-proton scenario a GeV counterpart to
the TeV emission is expected with comparable luminosity to the TeV
emission and hence detectable with \emph{Fermi}-LAT, albeit with
potential different morphology due to energy-dependent diffusion,
motivating the study presented here.

\section{\emph{Fermi}-LAT data and analysis}\label{sec:fermi}

The Large Area Telescope (LAT) onboard the {\it Fermi} satellite is a
pair-conversion instrument, operating in the 30~MeV to 300~GeV energy
range. The point-spread-function (PSF) of the LAT varies with energy
and becomes less than $0.5^\circ$ above $\sim$3\,GeV
\citep{Fermi:Inst}. The data set analysed here comprises a total of
$\sim$4.5 years of observations from August 2008 until January
2013. Only photons with energies $\geq 3$\,GeV are used, greatly
reducing the impact of the Galactic diffuse emission on the analysis,
minimising the contribution of the bright GeV-detected pulsar \psr\ in
this region and allowing us to search for multiple, spatially
separated components. The {\it Fermi} Science Tools package
\texttt{v9r27p1} and instrument response functions
\texttt{P7Source\_V6} are used. Photons from within a $14^\circ \times
14^\circ$ region centred at the optical Wd~1 position are used in a
{\it binned} maximum likelihood analysis. Sources that are listed in
the {\it Fermi} two-year catalogue (2FGL) and lie within 15$^\circ$ of
Wd~1 are modeled. The flux normalisations of objects within
$3.5^\circ$ are left free in the fit, with all other parameters fixed
to their 2FGL values. The Galactic diffuse component is modelled using
the ring hybrid model \texttt{ gal\_2yearp7v6\_v0.fits} with free
normalisation, and the isotropic extragalactic emission and particle
background according to the tabulated spectrum of
\texttt{iso\_p7v6source.txt}.

\begin{figure*}
  \resizebox{\hsize}{!}{\includegraphics[]{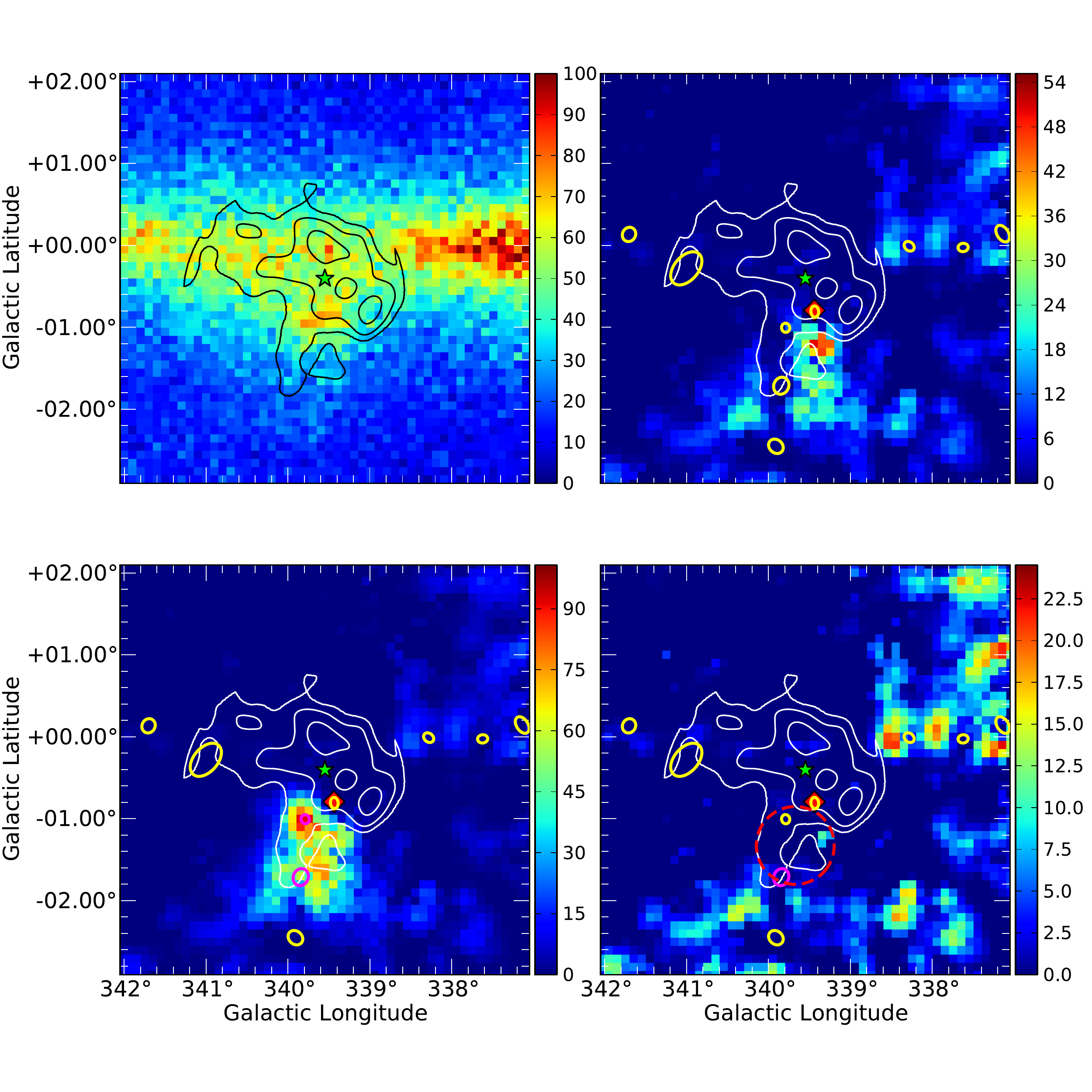}}
  \caption{{\it Fermi}-LAT counts map of the Wd~1 region between
    3\,GeV and 300\,GeV, with H.E.S.S. smoothed excess contours at the
    35\%, 55\%, and 85\% of the peak emission overlaid (top,
    left). The other three panels show residual TS maps in the same
    energy band. Yellow (magenta) ellipses indicate nearby 2FGL
    sources that are included (excluded) in the model fits. The green
    star denotes the Wd~1 stellar cluster position and the red diamond
    \psr. The residual TS maps show the 2FGL model (top,right), the
    2FGL without S1 and S2 model (bottom, left) and the best-fit model
    using \src\ instead of S1 (bottom, right). The red dashed line is
    the 1-$\sigma$ variance of \src.}
\label{fig1}
\end{figure*}

A search for diffuse HE \g-ray emission from the vicinity of Wd~1 is
performed for two different scenarios. First, we assume that all point
sources in the 2FGL catalogue {\it are not} associated to Wd~1 and
look for remaining \g-ray emission in the field-of-view. The residual
Test-Statistic (TS) map under this hypothesis is shown in
Figure~\ref{fig1} (top, right) and shows excess GeV emission south of
the stellar cluster. Second, we assume that 2FGL sources that lie
within the H.E.S.S. emission region {\it are} associated to Wd~1 and
therefore discard them from the model. The residual GeV emission after
excluding 2FGL~J1653.9$-$4627c (S1) and 2FGL~J1650.6$-$4603c (S2),
which are both flagged as potentially confused in the 2FGL catalogue,
from the source model is shown in the bottom, left panel of
Fig.~\ref{fig1}. The \g-ray excess is more pronounced than in the
first model, offset from Wd~1, and apparently extended with respect to
the LAT PSF. These TS maps imply that at least some of the emission
attributed to (in particular) S1 in the standard \emph{Fermi} model
may be due to a previously unidentified diffuse source in this region.

In the following we test different templates for diffuse \g-ray
emission. As significant TeV \g-rays are detected from the vicinity of
Wd~1, the VHE \g-ray excess map forms a natural template to test for
the GeV signal. The LAT data are not, however, well described by the
H.E.S.S. template and an upper limit on the HE \g-ray flux in the TeV
emission region can be obtained (see below). The second diffuse model
tested is a symmetric two-dimensional Gaussian. A grid-search in
position and extension is performed to find the maximum likelihood for
such a source model. The best-fit position of this template is at
\HMS{16}{51}{36}$\pm 24^{\mathrm{s}}$, Dec \DMS{-46}{21}{00}$\pm 5'$
(J2000), with a best-fit rms of $\sigma_{S} = (0.475 \pm 0.05)^\circ$.
For the rest of this work we will refer to this source as \src.
Including S2 in the source model gives a significantly better fit,
suggesting that S2 is a genuine additional point-like \g-ray source,
or that the Gaussian template is not an adequate description of the
diffuse source morphology.  The residual TS map of this model is shown
in the bottom, right panel of Fig~\ref{fig1} and illustrates that
there is relatively little residual emission from the vicinity of Wd~1
and the Galactic plane in this case. For this source model, the TS of
\src\ is 173, corresponding to a significance of 13.2$\sigma$
(pre-trials). As 4820 different combinations of source position and
$\sigma_{S}$ have been tested, the post-trials significance is $12.5$
standard deviations. The \g-ray spectrum between 3\,GeV and 300\,GeV
for \src\ is consistent with a power law in energy with \g-ray index
$\Gamma = 2.1 \pm 0.1$ and integral flux
$F(>$3\,GeV)=(4.7$\pm$0.5)$\times 10^{-9}$\flux. Note that
\citet{Neronov12} identified significant emission above 100~GeV from a
position consistent with this source.

\begin{figure}
  \centering
  \resizebox{1.07\hsize}{!}{\includegraphics[]{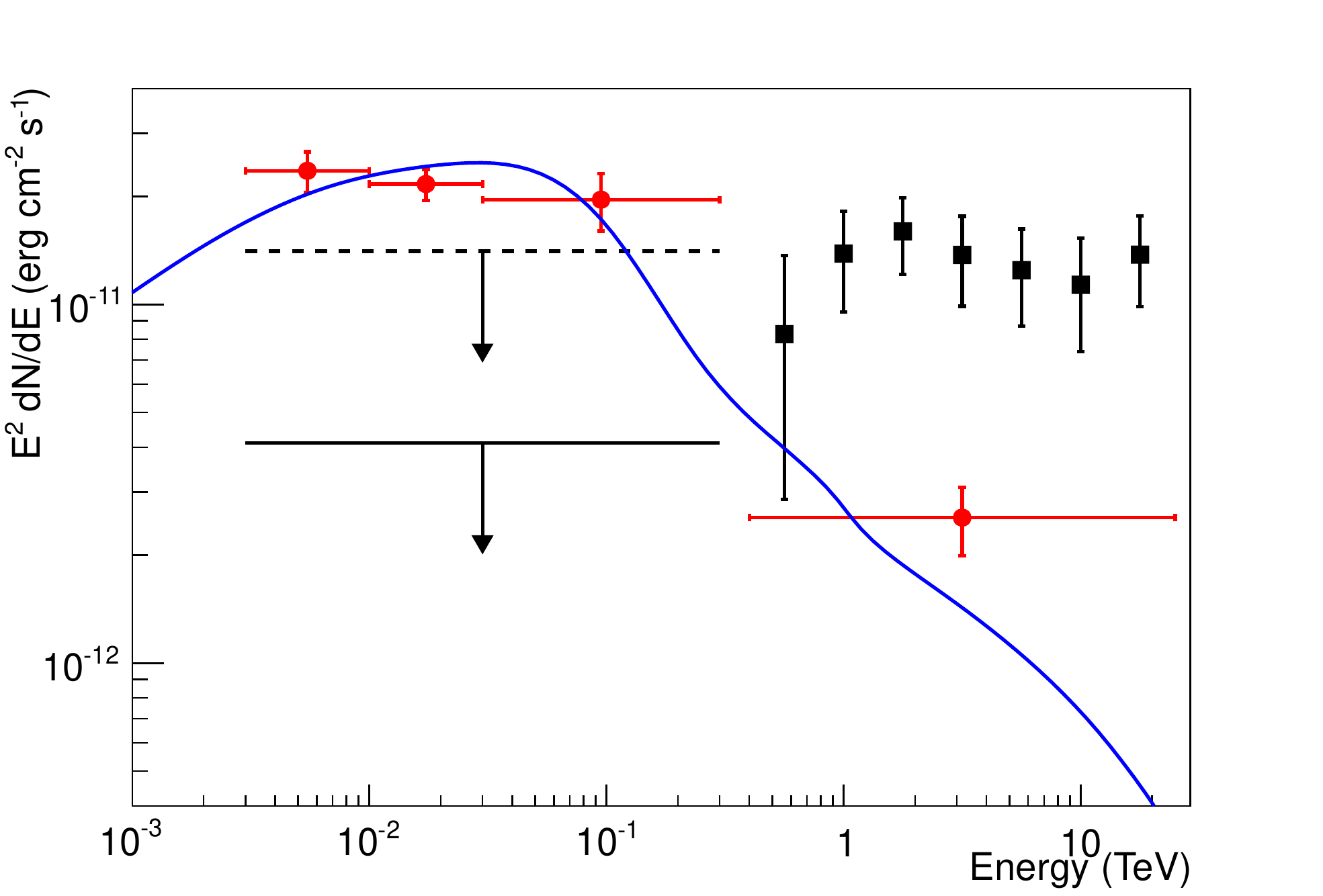}}
  \caption{\g-ray spectral energy distributions for two regions close
    to Wd~1. \src\, with TeV flux estimated from H.E.S.S. measurements
    as described in the text, is shown in red. The \g-ray spectrum of
    \hessj\ is shown in black. The upper limits on the GeV emission
    are obtained by excluding (dashed line) and including \src\ (solid
    line) in the model fit. The blue line is an illustrative model
    curve for IC emission in the PWN scenario discussed in
    Section~\ref{sec:pwn}.}
\label{fig2}
\end{figure}

Figure~\ref{fig2} shows the \g-ray spectrum of \src\ together with the
TeV data of \hessj. As \citet{HESS:Wd1_12} found no indications for
changes in source morphology with energy, we simply scale the total
TeV flux by the fractional VHE \g-ray excess to estimate the flux from
individual regions. The estimated VHE \g-ray flux from the template
region is based on the VHE flux integral within a 1-$\sigma$ radius
and corrected for missing flux assuming a Gaussian
profile. Fig.~\ref{fig2} also shows LAT upper limits for emission
following the H.E.S.S. template, with and without \src\ included in
the model. The log-likelihood and TS values for the different models
tested are listed in Table~\ref{tab1}. In an attempt to better
describe the diffuse GeV emission, we also tried wedge-shaped
Gaussian-like and top-hat templates. Such model fits, however, have
lower probabilities than the Gaussian template plus S2. The best-fit
\g-ray spectra for more complex templates are systematically steeper,
$\Gamma$$\simeq$2.3, which could either indicate energy-dependent
morphology of the diffuse source, or that S2 is indeed an unrelated
(and relatively steep spectrum) point source.

\begin{table}
  \centering
  \caption{Fit statistics for different source models. The change in
    significance of a model is given relative to the next best fit 
    model (i.e. the row above in the table). 2FGL~J1653.9$-$4627c is
    denoted as as S1 and 2FGL~J1650.6$-$4603c as S2. {\it 2FGL -
      S1 + Gauss}, for example, denotes the model which includes the full 2FGL
    catalogue of point sources {\it except} S1, plus a Gaussian
    template with properties described in the text.}
  \begin{tabular}{l c c}
    \hline
    Model & $\Delta\log{P}$ & $\sqrt{TS_{\rm model}}$ \\
    & & $\sigma$ \\\hline
    2FGL - S1,2& 0 & \\
    2FGL - S1,2 + H.E.S.S. & 4 & $+2.0$  \\
    2FGL & 75 & $+8.4$ \\
    2FGL - S1,2 + Gauss & 112 & $+6.1$  \\
    2FGL - S1 + Gauss & 125 & $+3.6$  \\ \hline
    \end{tabular}\label{tab1}
\end{table}

Wd~1 harbours at least 24 Wolf-Rayet stars, two LBV stars and many
super- and hyper-giants with a high binary fraction \citep[see][and
references therein]{Lim13} of which some show non-thermal emission
\citep{Dougherty10}. As one Colliding Wind Binary (CWB) in the Galaxy
is known to emit \g-ray emission, namely $\eta$~Car
\citep[e.g.][]{Reitberger12}, such emission might be expected from
Wd~1. Adding a point-like source at the position of the cluster to the
source model does not, however, significantly improve the fit. The
$2\sigma$ upper limit on the (3$-$300)\,GeV cluster \g-ray flux,
assuming a \g-ray spectral index of $\Gamma$=$-$2.0 is found to be
$F_{95}$(3$-$300$\,{\rm GeV}) <$3.2$\times
10^{-10}$\,ph\,cm$^{-2}$\,s$^{-1}$. The energy output of Wd~1 in
$\gamma$-ray emission from $(3-300)$\,GeV is
$<$1.5$\times10^{34}$\,erg\,s$^{-1}$. This is a factor of $\sim$4
lower than that of $\eta$~Car (6$\times10^{34}$\,erg\,s$^{-1}$),
tightly constraining the level of particle acceleration in CWBs in
this system. There are two known LBV stars in Wd~1 \citep[W9 and W243,
e.g.][]{Lim13}, of which W9 had a mass-loss rate in the recent past
comparable to the current mass-loss rate of $\eta$~Car
\citep{Dougherty10}. The $\gamma$-ray limit therefore implies that
either one or both of the LBV stars in Wd~1 are not binary systems,
or, if they are, that the wind power is much lower then in $\eta$~Car
or particle acceleration is for some reason much less efficient.

\section{Discussion}\label{sec:disc}

The dramatically different morphology present in the emission of this
region in the GeV and TeV bands implies that either these two sources
are unrelated, or that particle transport is playing an important
role. The presence of both a classical young pulsar and a magnetar
motivates an exploration of the scenario where the $\gamma$-ray
emission is dominated by an (electron-accelerating) PWN. Equally, the
presence of these stellar remnants implies supernova explosions in the
recent past ($\lesssim10^{5}$ years ago), which plausibly accelerated
protons and nuclei that are now interacting to produce the observed
\g-ray emission. Here we discuss these alternatives in turn.

\subsection{PWN scenario}\label{sec:pwn}

PWNe form the majority of identified Galactic TeV sources and a small
number of GeV associations. The efficient acceleration of particles in
PWN is observationally well established with acceleration of
e$^+$/e$^-$ pairs to PeV energies at the termination shock of the
relativistic pulsar wind required to explain the radio to VHE \g-ray
emission of these objects. The central energy source is a rotating
neutron star that converts rotational energy into high-energy
particles and subsequently into non-thermal radiation.  The energy
input to the PWN is a function of time and determined by the spin-down
luminosity $\dot{E}(t)$ of the pulsar:
\begin{equation}\label{eq1}
  \dot{E}(t) = \dot{E}_0/(1+t/\tau)^p,~\mathrm{with}~\tau =
  P_0/\dot{P}_0(n-1).
\end{equation}
Here, $\tau$ is the characteristic spin-down time, $P_0$ the pulsar
birth period and $\dot{P_0}$ the first derivative of $P_0$. The index
$p$ is defined as $p = (n + 1)/(n - 1)$ and $n$ is the braking
index. The injection of energetic particles can be approximated
assuming a constant fraction $\epsilon$ of $\dot{E}$ is converted to
relativistic electrons and positrons.

\psr\ is spatially coincident with part of the H.E.S.S. emission and
considered as possible counterpart to the VHE signal. The distance of
\psr\ as derived from dispersion measurements is 5.7\,kpc, with an
error of $\sim$30\% \citep{Kramer03}, and therefore consistent with
the estimated Wd~1 distance. In the following we will consider a
physical association of \psr\ and Wd~1. The age estimate of \psr\ of
$\sim$10$^{5}$ years implies that an associated PWN would be in an
evolved state and dominated by particles injected early on in its
evolution \citep[e.g.][]{deJager09}. Observations of \psr\ with the
Suzaku satellite revealed an extended source in the X-ray data, which
is interpreted as a PWN candidate \citep{Sakai13} and supports the
idea that some of the $\gamma$-ray emission might indeed originate in
the PWN. Given the proximity to Wd~1, radiation from member stars of
the cluster may contribute significantly to the target radiation field
for the inverse Compton (IC) process. Assuming a projected distance of
$\sim$0.4$^\circ$, we estimate that the contribution from Wd~1 is
comparable to the typical ISM radiation energy density of
$\sim$1\,eV/cm$^3$. Figure~\ref{fig2} shows the result of a
single-zone, time-dependent model, where particles are injected
according to Equation~\ref{eq1} \citep[e.g.][]{Funk07}. For the
measured current spin period $P_{\rm now}=165$\,ms, the inferred
characteristic age of 110\,kyr \citep{ATNF_Cat} and an assumed
conversion efficiency $\epsilon$ of 20\%, a pulsar birth period of
$P_{0}=21$\,ms is required to match the GeV flux. The radius of the
GeV emission region assuming a 4.3\,kpc distance is $\sim$40\,pc,
comparable to the extension of HESS~J1825$-$137 at these energies
\citep{Grondin11}. However, due to significant IC cooling on the
stellar cluster plus ISM plus CMB radiation fields, it is very
difficult to accomodate the TeV emission in the same scenario. This
emission could be attributed to (possibly multiple) additional (much
younger) PWNe, which have not so far been observed.

\subsection{SNR/proton scenario}

Given the presence of target material in the vicinity, and the likely
local acceleration of protons and nuclei in either supernova
explosions or cluster winds, a $\pi^{0}$-decay explanation for some or
all of the \g-ray emission is attractive.  Making quantitative
predictions in this scenario, with which to compare the observations,
is however rather difficult: the three dimensional distribution of
both the CRs and target material must be modelled or assumed. The
distribution of target material can be estimated from atomic hydrogen
(HI) and CO maps, but with considerable uncertainties on the
distribution of material along the line-of-sight and/or the presence
of dense or ionised material that could be missed. The distribution of
CRs is hard to predict due to: 1) uncertainties in the injection
spectrum, location and time for the CRs and 2) uncertainties on their
subsequent propagation. These uncertainties exist in essentially all
diffuse \g-ray sources, but are exacerbated in this case by the likely
presence of significant bulk motions (advection) as well as
energy-dependent and presumably environment dependent
diffusion. Nevertheless, we attempt here to construct a scenario which
is consistent with the observational data, as an illustration of the
kind of situation we may be dealing with here, which could be
constrained by a future precision \g-ray instrument such as CTA, and
to provide a reasonable estimate of the required energy input.

\begin{figure}
  \centering
  \resizebox{1.0\hsize}{!}{\includegraphics[]{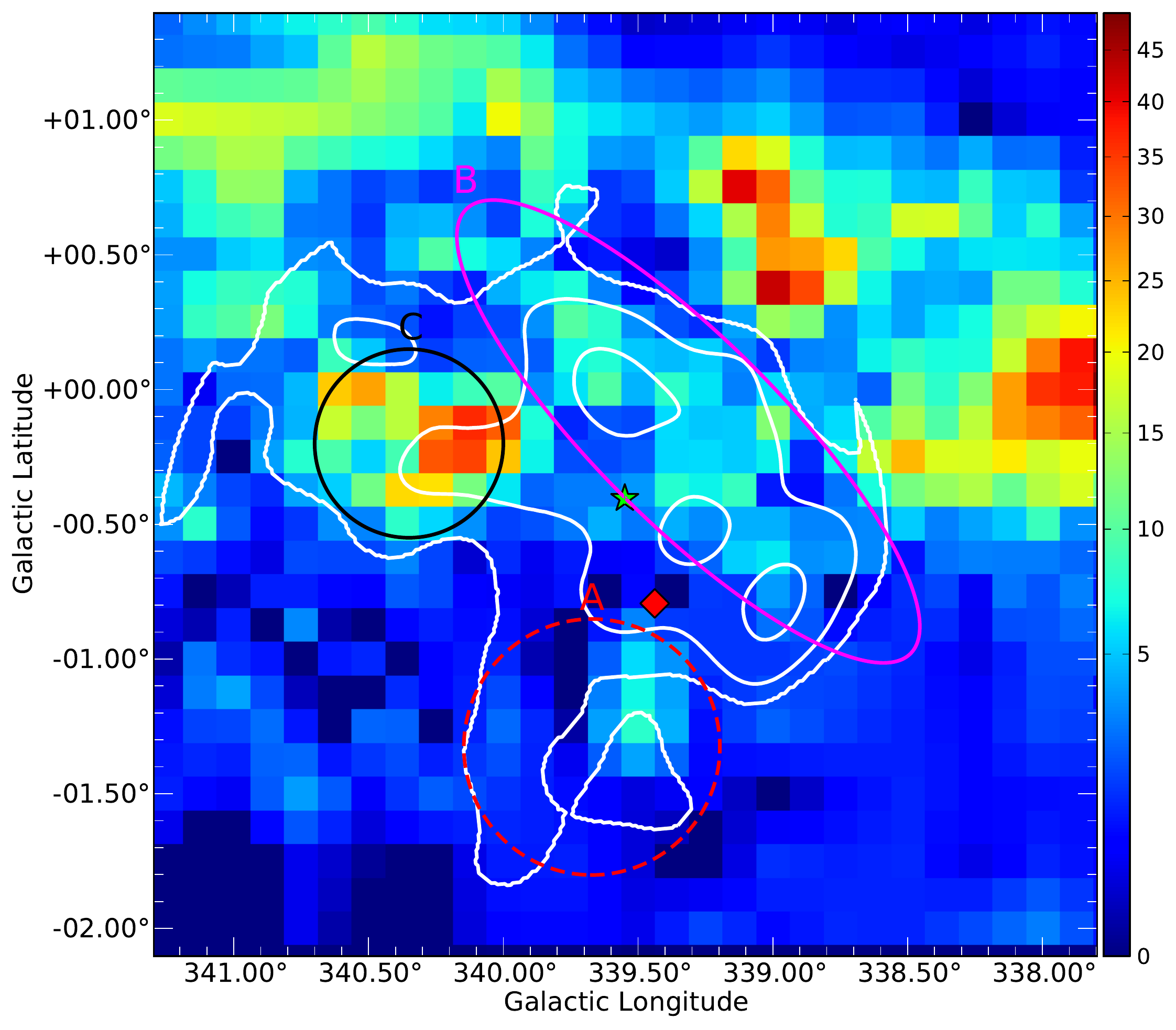}}
  \caption{$^{12}$CO map from \citet{Dame01}, integrated in the
    velocity range $-58.5$\,km\,s$^{-1}$ to
    $-52.0$\,km\,s$^{-1}$. Smoothed H.E.S.S. excess contours and
    markers are as described in Fig~\ref{fig1}. Regions A,B, and C are
    used to estimate the H$_2$ and HI masses and densities given in
    Table~\ref{tab2}.}
\label{fig3}
\end{figure}

Figure~\ref{fig3} shows the $^{12}$CO emission from this region,
tracing molecular hydrogen, with H.E.S.S. contours and best-fit
\emph{Fermi} ellipse superimposed. Three features/regions are apparent
in this map in the immediate vicinity of Wd~1:
\begin{itemize}
\item[\bf{A --}] a relatively low-density region to the south of Wd~1
\item[\bf{B --}] a molecular cloud complex coincident with the TeV peak
\item[\bf{C --}] the prominent star-forming region G340.2$-$0.2.
\end{itemize}
These regions have been investigated in terms of their HI emission as
well as CO, in order to provide density estimates. Integrating in the
velocity range $-58.5$\,km\,s$^{-1}$ to $-52.0$\,km\,s$^{-1}$ and
using conversion factors of $X_{\rm HI} = 1.8\times10^{18}$\,
cm$^{−2}$ / (K km s$^{-1}$) \citep{Yamamoto03} and $X_{\rm CO
  \rightarrow H_2} = 1.5\times10^{20}$\,cm$^{−2}$ / (K km s$^{-1}$)
\citep{Strong04}, we find total cloud masses between $\sim$(2-7)$
\times 10^5\,M_\odot$ (Table~\ref{tab2}). The velocity range has been
chosen following \citet{HESS:Wd1_12} and to account for possible local
gas motions of several km\,s$^{-1}$.

\begin{table}
  \centering
  \caption{Mass estimates for neutral and ionised hydrogen in the
    regions defined in Fig~\ref{fig3}. The average gas density
    estimates rely on the assumed three-dimensional structure and are
    therefore uncertain by a factor $\sim$2.}
  \begin{tabular}{c c c c}
    \hline
    Region & $M_{{\rm H}_2}$ & $M_{\rm HI}$ & $\bar{n}_{\rm H}$ \\
    & $10^{5} M_\odot$ & $10^{5}  M_\odot$ & cm$^{-3}$ \\\hline
    A & 0.9 & 1.0 & 5 \\
    B & 3.4 & 4.2 & 10  \\
    C & 4.0 & 1.2 & 35 \\ \hline
    \end{tabular}\label{tab2}
\end{table}


Given the possibility of different diffusion coefficients and even
different transport mechanisms in different regions, and the rather
complex geometry, we adopt a Monte-Carlo approach to the particle
transport. Three distinct zones are defined in which transport
properties and local density differ, corresponding approximately to
the regions A,B and C discussed above. Particles (2000 per energy bin,
with 20 energy bins per decade) are injected at the position of Wd~1
and followed for $t_s = 10^{5}$ years, or until they leave the region
of interest. The nominal $10^5$~year age is comparable to the
characteristic age of the magnetar ($1.7\times10^5$~years), but
significantly older than the estimated mean time between SN explosions
in the cluster ($\sim$10$^4$ years). In each 100-year time-step the
particles are considered to propagate by diffusion, resulting in a
random rms displacement of $\sqrt{6D(E, \mathbf{r}){\mathrm d}t}$,
where $D(E, \mathbf{r})$ is the energy and position (zone) dependent
diffusion coefficient, with a superimposed motion with fixed velocity
and direction for those particles in zone A, away from a point at the
northern edge of this zone. The energy dependence of the diffusion
coefficient is assumed to be $D(E)=D_{10}(E/10\,{\rm GeV})^{\delta}$,
with a (conventional) value of $\delta=0.6$ adopted. Smaller values of
$\delta$, however, make it hard to explain the energy-dependent
morphology observed. Different values of $D_{10}$ are tested in
different zones. At the end of the propagation simulation, particle
weights are multiplied by the local density and integrated in the
line-of-sight direction, to provide a map of integrated (CR density
$\times$ gas density) which can be used for the calculation of the
\g-ray flux. The parameterisations of \citet{Kamae06} are used to
calculate the \g-ray emission expected from each map pixel and finally
the \g-ray SED is integrated in each of regions A, B and C.

\begin{figure}
  \centering
  \resizebox{1.07\hsize}{!}{\includegraphics[]{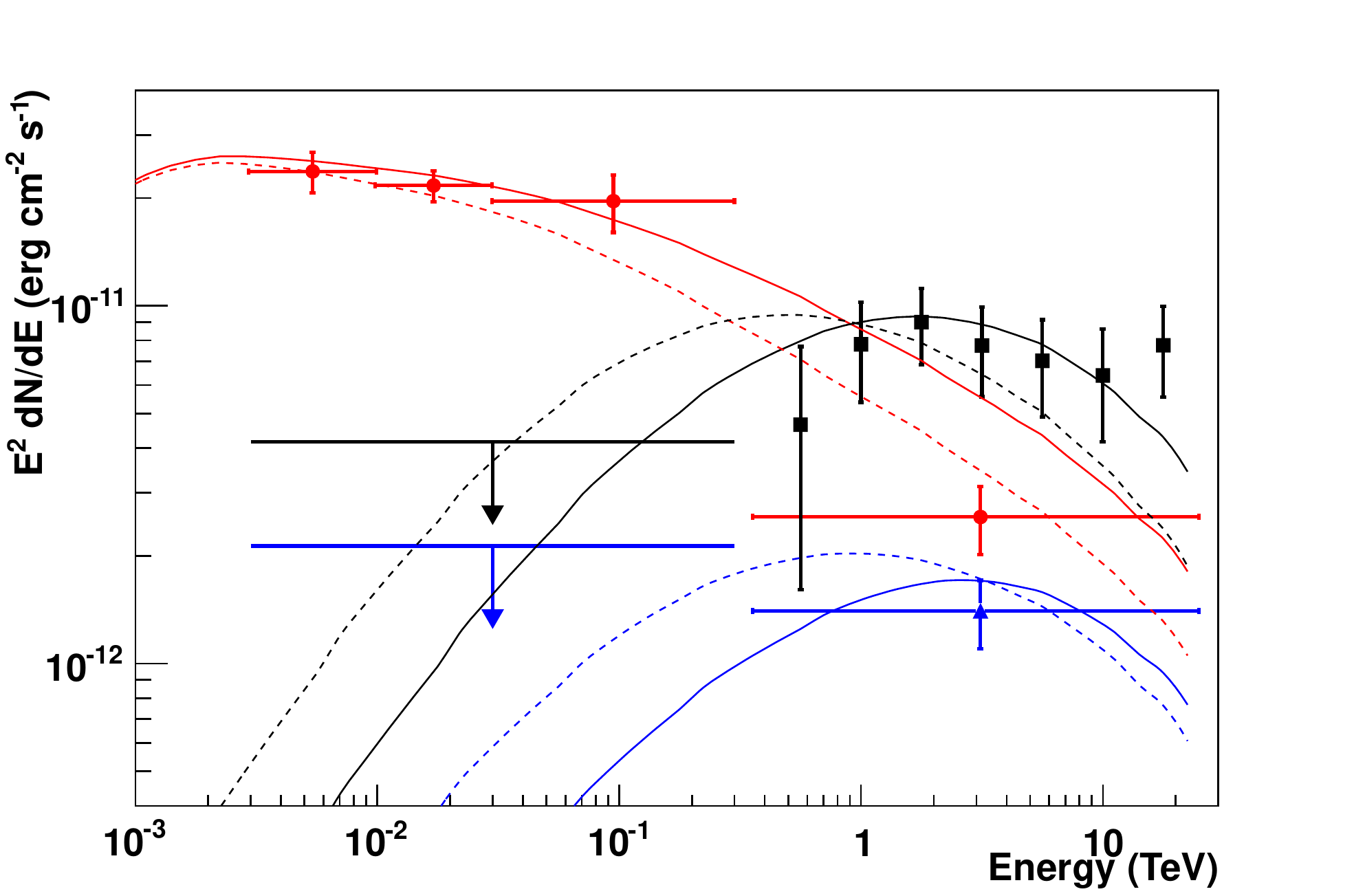}}
  \caption{Illustrative model curves for diffusing protons compared to
    the measured spectral energy information for regions A, B and C
    described in the text. Red points (circles) and curves correspond
    to region A, black (squares) to region B and blue (triangles) to
    region C. The black data points are taken from \citet{HESS:Wd1_12}
    and scaled down by the fractional excess contained in region
    A. The blue and red points are derived as described in Section
    \ref{sec:fermi}. The blue limit is derived by assuming an
    additional point source at the position of G340.2$-$0.2. Dashed
    model curves have diffusion coefficients $D_{10}$ of
    $2\times10^{25}$ cm$^{2}$ s$^{-1}$ (A and C) and
    $2/3\times10^{25}$ cm$^{2}$ s$^{-1}$ (B). The injection spectrum
    of relativistic protons follows a power law with index -2.05, with
    a total of $1.1\times10^{51}$ ergs injected. The solid curves are
    identical except for a decrease of $D_{10}$ in all regions by a
    factor 2.}
  \label{fig4}
\end{figure}

Figure~\ref{fig4} shows the result of two sets of such simulations,
which are consistent with the available measurements and upper limits
in the three zones. To provide a reasonable match to the observations
within this framework we require:
\begin{itemize}
\item Significantly slower diffusion than the typical values inferred
  for GCRs, and different diffusion speeds in Zones A and B. The
  conventional Galactic $D_{10}$ is 10$^{28}$ cm$^{2}$ s$^{-1}$.
  Values closer to $10^{25} (t_s/10^5\,{\rm yr})^{-1}$ cm$^{2}$
  s$^{-1}$ are required here, with factor $\sim$3 slower diffusion
  into the molecular clouds of Zone B. Such adjustments to the
  standard paradigm are uncomfortable, but the situation may overall
  be rather simular to that in W~28 \citep[e.g.][]{Fujita09, Ohira11,
    Li12}, particularly for a SN explosion $\sim$10$^{4}$ years ago.
\item A bulk flow with $v\sim$400~km/s towards the South (in Zone A)
  that corresponds to a $1^{\circ}$ offset from the cluster for material
  carried by the flow for $\sim$10$^{5}$ years. Such an outflow from the
  cluster seems plausible given the stellar population of Wd~1 and might be similar (in speed) to the winds observed in
  Starburst galaxies such as NGC\,253 \citep[e.g.][]{Zirakashvili06}.
\item An injection spectrum close to $E^{-2}$.
\item $\sim$10$^{51}$ erg injected in relativistic protons and nuclei.

\end{itemize}

Such a solution is certainly not unique, the problem being rather
unconstrained by the available, rather low resolution, \g-ray data,
and of course the time(s) at which particles are injected are highly
uncertain. The propagation scenario does however seem plausible. The
primary difficulty is the energy required in protons, which is close
to $10^{51}$ ergs for all models which provide reasonable agreement
with the data, constrained by the GeV flux and density in region
A. Multiple SNR and/or an extremely energetic event, both of which
seem plausible for Wd~1, would be required.

\section{Conclusions and Outlook}

We have established the existence of a new GeV source in the vicinity
of the massive stellar cluster Wd~1. The emission is extended in
nature and likely associated to the cluster either directly, via
collective wind effects, or indirectly via PWN(e) and/or SNR(s). A
single or multiple PWNe can naturally explain the GeV emission but
would leave the TeV emission unexplained. A scenario in which protons
from a single very energetic SNR or multiple SNRs diffuse away from
Wd~1 and interact with the environment has been investigated here and
can plausibly explain the \g-ray data with a required energy input of
$\sim$10$^{51}$\,erg. Wd~1 is the most massive stellar cluster in the
Galaxy hosting a magnetar and a rich population of evolved massive
stars. These unique characteristics imply that the SNR progenitor
stars had very high masses, i.e. $\gtrsim40\,M_\odot$, and that the
associated SN explosions may well have been very energetic. This in
turn could explain the large energy required in the scenario we employ
here. A detailed investigation of the underlying particle
acceleration, propagation and interaction processes, however, is
limited by the low resolution of \g-ray and CO data. To study this
complex and important region in more detail, requires high resolution,
more sensitive \g-ray instruments such as CTA as well as spatial X-ray
coverage and high-resolution radio data. Especially CTA with its wider
field-of-view and factor 5 better PSF will be crucial to study the
connection between the \emph{Fermi} and H.E.S.S. emission.

\section*{Acknowledgements}
The authors would like to thank the anonymous referee for his useful
comments and Thierry Montmerle for fruitful discussions. This work has
made use of public \emph{Fermi} data and Science Tools provided by the
Fermi Science Support Centre. S.O. acknowledges support from the
Humboldt foundation by a Feodor-Lynen fellowship.

\bibliographystyle{mn2e_williams}
\bibliography{Wd1_Fermi}
\label{lastpage}

\end{document}